\input harvmac
\input epsf
\def\gsim{{~\raise.15em\hbox{$>$}\kern-.85em
          \lower.35em\hbox{$\sim$}~}}
\def\lsim{{~\raise.15em\hbox{$<$}\kern-.85em
          \lower.35em\hbox{$\sim$}~}} 

\def\ra{\rightarrow}
\def\barrho{\bar\rho}
\def\bareta{\bar\eta}
\def\epsK{\varepsilon_K}
\def\barB{\overline{B^0}}
\def\BBbar{B^0-\barB}
\def\KKbar{K^0-\overline{K^0}}
\def\BBsbar{B_s-\overline{B_s}}
\def\apks{a_{\psi K_S}}

\def\Im{{\cal I}m}
\def\epe{\varepsilon^\prime/\varepsilon}

\noblackbox
\baselineskip 18pt plus 2pt minus 2pt
\Title{\vbox{\baselineskip12pt
\hbox{hep-ph/0008009}
\hbox{WIS/12/00-Aug-DPP}
\vskip -1.5truecm
}}
{\vbox{ 
\centerline{Implications of a Small CP Asymmetry in $B\ra\psi K_S$}
} }
\vskip -1truecm
\centerline{Galit Eyal$^a$, Yosef Nir$^{b}$ and Gilad Perez$^b$}
\medskip
\centerline{\it $^a$Research Department, Camelot Information 
Technologies Ltd., Matam, Haifa 31905, Israel}
\smallskip
\centerline{\it $^b$Department of Particle Physics,
Weizmann Institute of Science, Rehovot 76100, Israel}
\smallskip
\bigskip

\baselineskip 18pt
\noindent
New measurements of the CP asymmetry in $B\ra\psi K_S$, $\apks$, 
by the BABAR and BELLE collaborations are consistent with the
Standard Model prediction for $\sin2\beta$. These measurements,
however, leave open the possibility that $\apks$ is well below
the Standard Model prediction. We identify deviations from
the `reasonable ranges' of hadronic parameters that can lead to
low values of $\sin2\beta$. New physics, mainly in $\BBbar$ mixing
and/or $\KKbar$ mixing, can explain low values of $\apks$ in two ways:
either by allowing for values of $\sin2\beta$ below the Standard
Model prediction or by modifying the relation between $\apks$ and
$\sin2\beta$. 

\vfill

\Date{8/00}

\nref\babar{D. Hitlin, BaBar collaboration, plenary talk
in ICHEP 2000 (Osaka, Japan, July 31, 2000), SLAC-PUB-8540.}%
\nref\belle{H. Aihara, Belle collaboration, plenary talk
in ICHEP 2000 (Osaka, Japan, July 31, 2000).}%
\nref\cdf{T. Affolder {\it et al.}, CDF collaboration,
Phys. Rev. D61 (2000) 072005 [hep-ex/9909003].}%
\nref\opal{K. Ackerstaff {\it et al.}, OPAL collaboration,
 Eur. Phys. J. C5 (1998) 379 [hep-ex/9801022].}%
\nref\cdfo{F. Abe {\it et al.}, CDF collaboration,
 Phys. Rev. Lett. 81 (1998) 5513 [hep-ex/9806025].}%
\nref\aleph{ALEPH collaboration, ALEPH-99-099 CONF-99-54 (1999).}%
\newsec{Introduction}
Recent studies of the time dependent $B\ra\psi K_S$ decay rates
by the BaBar \babar\ and BELLE \belle\ collaborations have provided
highly interesting results concerning CP violation in $B$ decays.
When combined with a previous CDF result \cdf, the average
measured CP asymmetry is
\eqn\aveapks{\apks=0.42\pm0.24,}
where $\apks$ is defined via
\eqn\defapks{{\Gamma[\barB(t)\ra\psi K_S]-\Gamma[B^0(t)\ra\psi K_S]
\over \Gamma[\barB(t)\ra\psi K_S]+\Gamma[B^0(t)\ra\psi K_S]}=
\apks\sin(\Delta m_B t).}
(For previous measurements, see \refs{\opal-\aleph}.)
Within the Standard Model, $\apks$ is related to the angle $\beta$
of the unitarity triangle,
\eqn\defbet{\apks=\sin2\beta,\ \ \ \beta\equiv\arg\left[-
{V_{cd}V_{cb}^*\over V_{td}V_{tb}^*}\right].}
Based on the detrmination of the CKM parameters through measurements
of $|V_{ub}/V_{cb}|$, $\epsK$, $\Delta m_B$ and $\Delta m_{B_s}$,
the Standard Model (SM) prediction is
\eqn\numstb{0.59\lsim\sin2\beta\lsim0.82.}
Thus, the measurement of $\apks$ \aveapks\ is consistent with the 
SM prediction \numstb. This can be seen in Fig. 1 where the various
bounds are given in the plane of the two Wolfenstein parameters,
$(\barrho,\bareta)$. Yet, the allowed range in \aveapks\ leaves
open the possibility that $\apks$ is actually significantly smaller
than the SM prediction. It is this possibility that we investigate
in this work. For the sake of concreteness, we will assume that $\apks$ 
lies indeed within the 1$\sigma$ range of the BaBar measurement \babar,
\eqn\numapks{\apks=0.12\pm0.38.}
While negative values of $\apks$ are unlikely in view of the other
measurements, our main interest will be in the upper bound,
$\apks\lsim0.5$.

\centerline{\epsfbox{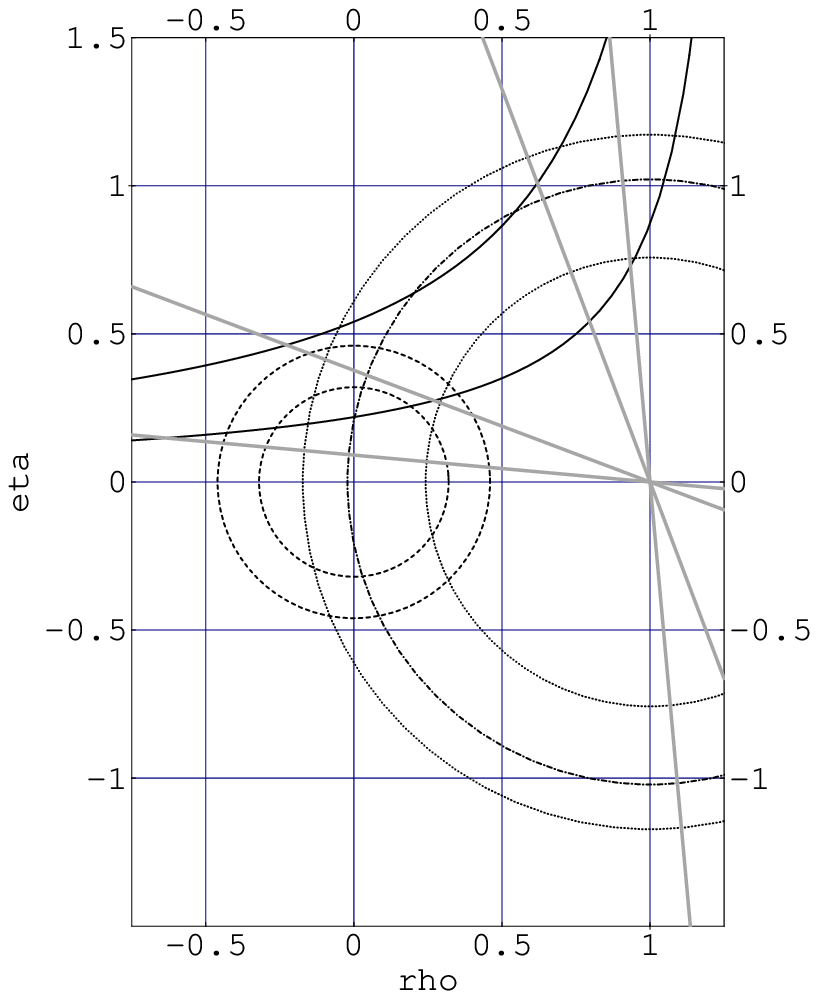}}

\noindent   
{\rm Figure 1. } Present constraints on the apex $(\barrho,\bareta)$
of the unitarity triangle: $|V_{ub}/V_{cb}|$ (dashed), $\Delta m_{B_d}$
(dotted), $\Delta m_{B_d}/\Delta m_{B_s}$ (dash-dotted), $\epsK$
(solid), and the average of the CDF, BELLE and BABAR measurements
of $\apks$ (thick grey lines).
\medskip

If, indeed, $\apks\lsim0.5$, there are two ways in which the
conflict with \numstb\ might be resolved:
\item{(i)} The SM is valid but one or more of the hadronic parameters 
which play a role in the analysis that leads to \numstb\ are outside
their `reasonable range'. We discuss this possibility in section 2.
\item{(ii)} New physics affects the CP asymmetry in $B\ra\psi K_S$
and/or some of the measurements that lead to \numstb. We discuss
this possibility in section 3.

We summarize our conclusions in section 4.

\nref\nirssi{Y. Nir, lectures given in the XXVII SLAC Summer Institute
on Particle Physics (SLAC, July 7-16, 1999), hep-ph/9911321.}%
\nref\babook{The BaBar Physics Book, eds. P. Harrison and H.R. Quinn,
 SLAC-R-504 (1998).}%
\nref\plas{S. Plaszczynski, invited talk at the XII Recontres
 de Physique de la Valee d'Aoste (La Thuile, March 1-7, 1998),
 hep-ph/9804330.}%
\nref\gnps{Y. Grossman, Y. Nir, S. Plaszczynski and M.-H. Schune,
 Nucl. Phys. B511 (1998) 69 [hep-ph/9709288].}%
\nref\PlSc{S. Plaszczynski and M.-H. Schune, talk at 8th International 
 Symposium on Heavy Flavor Physics (Southampton, 25-29 July, 1999),
 hep-ph/9911280.}%
\nref\burall{A.J. Buras, lectures given in the 14th Lake Louis
 Winter Institute (February 14-20, 1999), hep-ph/9905437.}%
\nref\AlLo{A. Ali and D. London, invited talk in the 3rd Workshop
on Physics and Detectors for Daphne (Frascati, November 16-19, 1999),
hep-ph/0002167.}%
\nref\BePe{S. Bergmann and G. Perez, hep-ph/0007170.}%
\nref\bususy{A.J. Buras {\it et al.}, hep-ph/0007313.}%
\newsec{Hadronic Uncertainties}
The computations that relate experimental observables to
CKM parameters suffer, in general, from theoretical uncertainties.
In very few cases, the calculation is made entirely in the
framework of a systematic expansion and it is possible to
reliably estimate the error that is induced by truncating the
expansion at a finite order. This is the case with the
relation between the observable $\apks$ and the CKM parameter
$\sin2\beta$: Within the SM, the relation \defbet\ holds to
an accuracy of better than one percent (for a review, see \nirssi).
Thus, if we assume that \numapks\ holds, we have
\eqn\acpcon{-0.26\lsim{2\bareta(1-\barrho)\over
\bareta^2+(1-\barrho)^2}\lsim+0.50.}

In most cases, however, the calculation involves models,
ansatze or, on occasion, educated guesses and there is no
easy way to estimate the errors that are involved. This is
the case with almost all the observables that are involved
in the prediction \numstb. We will follow the treatment of
this issue of ref. \babook. We will quote `reasonable ranges'
for the parameters that involve uncontrolled theoretical
uncertainties, and compare them to the values that are 
required for consistency with \numapks.

We emphasize that we use rather conservative CKM constraints, similar to 
those of refs. \refs{\babook-\bususy}. In particular, these ranges arise
when the BaBar method \refs{\babook-\PlSc} or a flat distribution
\refs{\burall-\bususy} are employed for theoretical errors. 
Studies that assume that theoretical errors have a gaussian 
distribution obtain stronger constraints.
Our input parameters are taken from \refs{\burall,\bususy}\
and the resulting constraints were calculated in \BePe. 

The $R_u$ parameter is determined from semileptonic charmless
$B$ decays:
\eqn\ruexp{R_u\equiv\sqrt{\bar\rho^2+\bar\eta^2}
={1\over\lambda}\left|{V_{ub}\over V_{cb}}\right|.}
Using 
\eqn\VubVcb{\eqalign{|V_{ub}|\ =&\ (3.56\pm0.56)\times10^{-3},\cr
|V_{cb}|\ =&\ (4.0\pm0.2)\times10^{-2},\cr}}
we obtain
\eqn\Rsubu{\sqrt{\bar\rho^2+\bar\eta^2}=0.39\pm0.07.}
Most of the uncertainty comes from hadronic modelling and
therefore should be viewed as a `reasonable range' rather
than a 1$\sigma$ error.

The $R_t$ parameter is given by
\eqn\rtexp{R_t\equiv\sqrt{(1-\bar\rho)^2+\bar\eta^2}
={1\over\lambda}\left|{V_{td}\over V_{ts}}\right|,}
where $V_{td}$ is determined from $\Delta m_B$ \burall,
\eqn\dmbvtd{{|V_{td}|\over8.8\times10^{-3}}=
{0.2\ GeV\over\sqrt{B_{B_d}}f_{B_d}}\left({170\ GeV\over m_t}\right)^{0.76}
\left({\Delta m_B\over0.5\ ps^{-1}}\right)^{1/2}
\left({0.55\over\eta_B}\right)^{1/2},}
or from $\Delta m_{B_d}/\Delta m_{B_s}$:
\eqn\ratds{\left|{V_{td}\over V_{ts}}\right|=\xi
\left({m_{B_s}\over m_{B_d}}\right)^{1/2}
\left({\Delta m_{B_d}\over\Delta m_{B_s}}\right)^{1/2}.}
Using
\eqn\VtdVts{\eqalign{\sqrt{B_{B_d}}f_{B_d}\ =&\ 0.20\pm0.04\ GeV,\cr
m_t\ =&\ 165\pm5\ GeV,\cr
\Delta m_B\ =&\ 0.471\pm0.016\ ps^{-1},\cr
\Delta m_{B_s}\ \geq&\ 14.6\ ps^{-1},\cr
\xi\ =&\ 1.14\pm0.08,\cr}}
we obtain
\eqn\Rsubt{\sqrt{(1-\bar\rho)^2+\bar\eta^2}=0.98^{+0.04}_{-0.22}.}
There are two theoretical parameters in \VtdVts.
The uncertainty in $\sqrt{B_{B_d}}f_{B_d}$ comes from systematic
errors in lattice calculations and should be viewed as a `reasonable
range'. On the other hand, the deviation from $\xi=1$ is a
correction to the SU(3) limit and believed to be under much better control.
 
The $\epsK$ constraint reads:
\eqn\defepsk{\bar\eta\left[(1-\bar\rho)A^2\eta_2 F_{tt}+P_c(\epsilon)
\right]A^2\hat B_K=0.226.}
Using \burall\
\eqn\FPetc{\eqalign{
A\ =&\ 0.826\pm0.041,\cr
\eta_2\ =&\ 0.57\pm0.01,\cr
F_{tt}\ =&\ 2.46(m_t/170\ GeV)^{1.52},\cr
P_c(\epsilon)\ =&\ 0.31\pm0.05,\cr
\hat B_K\ =&\ 0.80\pm0.15,\cr}}
we obtain:
\eqn\epscon{\bar\eta\left[(1-\bar\rho)(0.91^{+0.16}_{-0.14})+(0.31\pm0.05)
\right]=0.41^{+0.15}_{-0.10}.}
The main contribution to the uncertainty on the right hand side of
eq. \epscon\ comes from the theoretically calculated $\hat B_K$
and, again, represents a reasonable range.

\centerline{\epsfbox{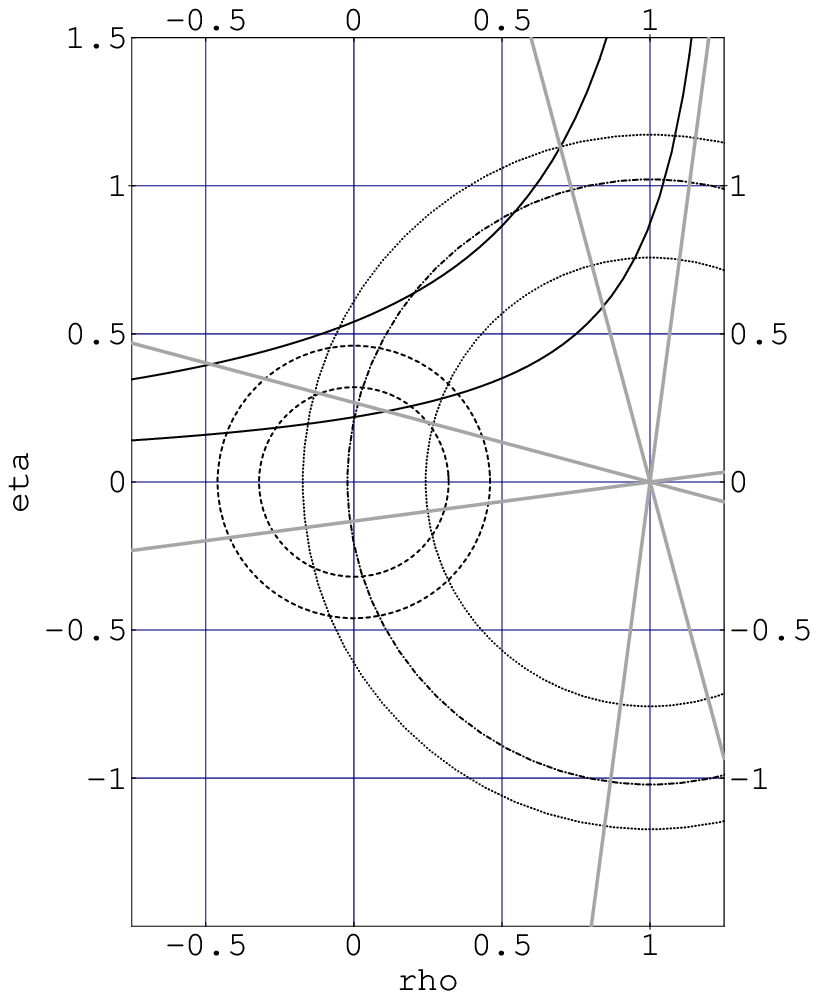}}

\noindent
{\rm Figure 2. } Present constraints on the apex $(\barrho,\bareta)$
of the unitarity triangle: $|V_{ub}/V_{cb}|$ (dashed), $\Delta m_{B_d}$
(dotted), $\Delta m_{B_d}/\Delta m_{B_s}$ (dash-dotted), $\epsK$
(solid), and the BABAR measurement of $\apks$ (thick grey lines).
\medskip

The four constraints \acpcon, \Rsubu, \Rsubt\ and \epscon\ are given
in Fig. 2.  There is no overlap between the four allowed ranges.
The last three together form an allowed range in the 
$\barrho-\bareta$ plane which can be translated to the
constraint on $\sin2\beta$ given in eq. \numstb. One lesson
which should be clear from our discussion here is that there
are several hadronic parameters that enter into this constraint.
In particular, for $R_u$, $\sqrt{B_{B_d}}f_{B_d}$ and $\hat B_K$
we are only able to quote reasonable ranges. We now ask what values
of these parameters will allow $\sin2\beta\lsim0.5$, consistently
with \numapks.

To answer this question, we study whether the discrepancy
between \numstb\ and \numapks\ can be explained by a single
theoretical parameter that is outside of its reasonable range.

(i) Let us assume that the $R_t$ constraint \Rsubt\ and the $\epsK$
constraint \epscon\ hold, but allow a failure of the $R_u$ constraint 
\Rsubu. Indeed, under such assumptions, there is an allowed region
around $(\barrho,\bareta)=(0.01,0.26)$. It requires, however, that
\eqn\vioru{R_u\lsim0.27,}
which is below its reasonable range in \Rsubu. In other words,
if the inconsistency comes from the hadronic modelling of charmless
semileptonic $B$ decays, the failure of these models should be such
that $|V_{ub}|$ is about 30\% lower than the presently most favorable value.

There is a second allowed region, around $(\barrho,\bareta)=(0.8,0.8)$.
Here, however, $R_u\gsim1.1$ is required. This value is about three
times larger than the best present estimate.
We find such a situation very unlikely.

(ii) Let us assume that the $R_u$ constraint \Rsubu\ and the $\epsK$
constraint \epscon\ hold, but allow a failure of the $R_t$ constraint 
\Rsubt. Under such assumptions, there is an allowed region
around $(\barrho,\bareta)=(-0.3,0.3)$. It requires, however, that
\eqn\viort{\xi\gsim1.4,}
which is well above the $1\sigma$ range in \VtdVts.
Note, however, that $\sqrt{B_{B_d}}f_{B_d}$ could be within its
reasonable range (close to the lower bound).

(iii) Let us assume that the $R_u$ constraint \Rsubu\ and the $R_t$
constraint \Rsubt\ hold, but allow a failure of the $\epsK$ constraint 
\epscon. There is a small viable region around 
$(\barrho,\bareta)=(0.25,0.20)$. It requires, however, that
\eqn\viobk{\hat B_K\gsim1.3,}
which is well above its reasonable range in \FPetc. 

To summarize: assuming that the CP asymmetry in $B\ra\psi K_S$
is within the $1\sigma$ range of BaBar measurement,
the SM could still be valid if some of the hadronic parameters
are outside of their `reasonable' ranges. If the apparent discrepancy 
is related to an error in the theoretical estimate of just one parameter,
then it requires either a small value of $|V_{ub}|$, or a large value
of $\xi$ or a large value of $\hat B_K$. The first of these,
$|V_{ub}/V_{cb}|\lsim0.06$, is perhaps the least unlikely deviation
from our `reasonable ranges.'
 
\newsec{New Physics}
New physics can explain an inconsistency of $\apks$ 
measurement with the SM predictions. It can do so provided
that it contributes significantly either to $\BBbar$
mixing or to the CP violating part of $\KKbar$ mixing
or to both. In this section we examine each of these possibilities. 

It is also possible, in principle, that the discrepancy is explained 
by a new contribution to $b\ra u\ell\nu$ decays or to 
$b\ra c\bar cs$ decays. We find it unlikely, however, that these SM 
tree level decays are significantly affected by new physics. 
In particular, we assume that the charmless semileptonic $B$ decays 
provide a valid measurement of $|V_{ub}|$ and, consequently,
\Rsubu\ holds.

\nref\SoWo{J.M. Soares and L. Wolfenstein,
 Phys. Rev. D47 (1993) 1021.}%
\nref\DDO{N.G. Deshpande, B. Dutta and S. Oh,
 Phys. Rev. Lett. 77 (1996) 4499 [hep-ph/9608231].}%
\nref\SiWo{J.P. Silva and L. Wolfenstein, 
 Phys. Rev. D55 (1997) 5331 [hep-ph/9610208].}%
\nref\CKLN{A.G. Cohen, D.B. Kaplan, F. Lepeintre and A.E. Nelson,
 Phys. Rev. Lett. 78 (1997) 2300 [hep-ph/9610252].}%
\nref\GNW{Y. Grossman, Y. Nir and M.P. Worah, 
 Phys. Lett. B407 (1997) 307 [hep-ph/9704287].}%
\subsec{$\BBbar$ mixing}
The effects of new physics on $\BBbar$ mixing can be described
by a positive dimensionless parameter $r_d^2$ and 
a phase $2\theta_d$ \refs{\SoWo-\GNW}:
\eqn\motnp{M_{12}=r_d^2 e^{2i\theta_d}M_{12}^{\rm SM}.}
Here $M_{12}$ ($M_{12}^{\rm SM}$) is the full (SM) 
$\BBbar$ mixing amplitude. 

If the new physics modifies the phase of the mixing amplitude,
$2\theta_d\neq0$, then the CP asymmetry in $B\ra\psi K_S$ is 
modified. Instead of eq. \defbet\ we now have:
\eqn\apksnp{\apks=\sin2(\beta+\theta_d).}
If the new physics modifies the magnitude of the mixing 
amplitude, $r_d^2\neq0$, then $\Delta m_B$ is modified.
In eq. \dmbvtd, we should replace $|V_{td}|$ with $r_d|V_{td}|$.
In addition, if the new physics modifies the $\BBsbar$ mixing 
amplitude, and we parametrize this modification
with corresponding parameters $r_s^2$ and $2\theta_s$, then instead
of eq. \ratds\ we now have:  
\eqn\ratdsnp{\left|{V_{td}\over V_{ts}}\right|=\xi\ {r_s\over r_d}
\left({m_{B_s}\over m_{B_d}}\right)^{1/2}
\left({\Delta m_{B_d}\over\Delta m_{B_s}}\right)^{1/2}.} 

If there is no new physics in the $R_u$ and $\epsK$ constraints,
then the following bounds hold:
\eqn\vtdbnpb{\eqalign{0.7\ \lsim R_t&\ \lsim1.4,\cr
12^o\ \lsim2\beta&\ \lsim55^o.\cr}}
Furthermore, there are correlations between the allowed ranges.
To achieve consistency with the measurements concerning $\BBbar$
mixing, it is required that either (i) $r_d/r_s\neq1$,
or (ii) $2\theta_d\neq0$ or (iii) both. 

\item{(i)} Consider models where there is a new contribution to 
both $\BBbar$ mixing and $\BBsbar$, but at least the first of these 
carries the same phase as the Standard Model contribution. 
With $2\theta_d=0$, we need 
\eqn\rnpb{0.5\lsim r_d\lsim1,\ \ \ r_s/r_d\gsim1.1.}
\item{(ii)} Consider the case where there
is a new CP violating contribution to $\BBbar$ mixing which, however,
keeps the magnitude of the mixing amplitude unchanged. We know of
no mechanism that would predict such a scenario, which requires
a relation between the magnitude and the phase of $M_{12}^{\rm NP}$,
the new physics contribution to $M_{12}$. We view such a situation 
as a rather unlikely accident. For completeness, we quote the required
$\theta_d$ range for $r_d^2=r_s^2=1$: 
\eqn\tnpb{2\theta_d\in\{95^o,160^o\}\ {\rm or}\ \{290^o,355^o\}.}
\item{(iii)} With generic new contributions to $\BBbar$ and $\BBsbar$
mixing, we can accommodate the experimental measurements with
\eqn\rtnpb{\eqalign{
0.5\lsim r_d&\lsim1.7,\cr
r_s/r_d&\gsim0.7,\cr
2\theta_d&\in\{0^o,18^o\}\ {\rm or}\ \{95^o,183^o\}\ {\rm or}\ 
\{290^o,360^o\}.\cr}}

Two points are in order:

1. If the new physics contribution carries no new phase, then
it must be flavor violating in the sense that $r_s\neq r_d$.
In other words, the flavor structure should be different
from the CKM one.

%

2. The contribution of the new physics to the mixing amplitude
cannot be much smaller than the Standard Model one. To be
more quantitative on this point, it is convenient to replace
the $(r_d^2,2\theta_d)$ parametrization of eq. \motnp\ with
a different parametrization defined as follows:
\eqn\defhsig{M_{12}^{\rm NP}=he^{i\sigma}M_{12}^{\rm SM},}
that is, $r_d^2e^{2i\theta_d}=1+he^{i\sigma}$. The $h$ parameter
gives the ratio between the size of the new physics contribution
and the SM one. 
We numerically scanned the allowed region for $h$ and $\sigma$. 
(We take in this scan $r_s=1$, so that the new physics effects
are restricted to $\BBbar$ mixing.) We find that
\eqn\npvssm{h\equiv|M_{12}^{\rm NP}/M_{12}^{\rm SM}|\gsim0.1.}

\nref\bggjs{A.J. Buras {\it et al.}, hep-ph/0007085.}%
\subsec{$\KKbar$ mixing}
A value of $\apks$ below the SM prediction can arise
even if there is no new physics in $\BBbar$ mixing and in
$b\ra c\bar cs$ decay, the two processes that are relevant 
to the CP asymmetry in $B\ra\psi K_S$.
The explanation must then be related to processes (other
than $\Delta m_B$) that play a role in constraining the 
$\sin2\beta$ range. The prime suspect is CP violation in the
neutral kaon system, that is $\epsK$.

If $\epsK$ is not fully explained by the SM, then the
$\epsK$ constraint, eq. \defepsk, does not hold. 
On the other hand, with no significant new physics
contributions to the mixing and the relevant decays of the 
$B$-mesons, the $R_u$, $\Delta m_{B_q}$ and $\apks$ constraints
hold. As explained in the previous section, in such a case
there is a small region around $(\barrho,\bareta)=(0.25,0.20)$
that is marginally consistent with all of these constraints when 
the hadronic parameters reside within our `reasonable ranges.'
In this region, the new physics has to add up constructively
to the SM contribution to $\epsK$, with 
\eqn\npeps{\Im\ M_{12}^{\rm NP}(K)/\Im\ M_{12}^{\rm SM}(K)\gsim0.3.}

The situation that there is no new physics in $\BBbar$ mixing
and $\BBsbar$ mixing has also interesting implications for
$\epe$. Within the SM, it is not easy to explain the measured
value of $\epe$, and various hadronic parameters have to assume
values outside their reasonable ranges. In the scenario discussed
in this section, the puzzle becomes even stronger. The combination
of the constraints on $\sin2\beta$ and on $R_t$ implies that
$\bareta$ is rather small. Even if we allow mild deviations from our 
range for either $|V_{ub}/V_{cb}|$ or $\sqrt{B_{B_d}}f_{B_d}$,
the remaining $\Delta m_{B_s}/\Delta m_{B_d}$ and $\apks$
constraints are enough to give $\bareta\lsim0.25$. Thus, even
if the new physics contribution to $\epsK$ can be rather small,
it should also add up constructively to the SM contribution to $\epe$.
 
Note that the situation where the $R_u$, 
$\Delta m_{B_s}/\Delta m_{B_d}$ and $\apks$ constraints
are valid but the $\epsK$ and $\Delta m_{B_d}$ are not
arises in models of new physics where all flavor violation 
and CP violation are described by the CKM matrix. This class of models
was defined and analyzed in ref. \bggjs. We find that, if \numapks\
holds, the new physics should probably play a role in both
$\epsK$ and $\epe$.

\nref\ben{G. Barenboim, G. Eyal and Y. Nir,
 Phys. Rev. Lett. 83 (1999) 4486 [hep-ph/9905397].}%
\nref\EyNi{G. Eyal and Y. Nir, JHEP 09 (1999) 013 [hep-ph/9908296].}%
\subsec{Neutral meson mixing}
In a large class of models, there could be significant contributions
to both $\BBbar$ mixing and $\KKbar$ mixing. However, $b\ra u\ell\nu$
decays are dominated by the $W$-mediated tree level decay.
The implications of measurements of $\apks$ in such a framework
were recenty investigated in refs. \refs{\ben,\EyNi}.
In such a framework, only the $R_u$ constraint of eq. \Rsubu\ holds.
In particular, the upper bound $R_u\lsim0.46$ gives the following
constraint on $\beta$:
\eqn\sintb{|\sin2\beta|\lsim0.82,\ \ \ \cos2\beta\gsim0.58.}
The assumed value of $\apks$ can be accommodated with
$-0.940\lsim\sin2\theta_d\lsim+0.996$.
(The bounds on $r_d^2$ are similar to \ben.)

\nref\NiSi{Y. Nir and D. Silverman, Nucl. Phys. B345 (1990) 301.}%
\nref\gnr{Y. Grossman, Y. Nir and R. Rattazzi, in {\it Heavy Flavours 
 II}, eds. A.J. Buras and M. Lindner, World Scientific Publishing Co., 
 (Singapore, 1997) [hep-ph/9701231].}%
\nref\dgh{M. Dugan, B. Grinstein and L.J. Hall, 
 Nucl. Phys. B255 (1985) 413.}%
\nref\BuSi{A.J. Buras and L. Silvestrini,
 Nucl. Phys. B546 (1999) 299 [hep-ph/9811471].}%
\nref\CoIs{G. Colangelo and G. Isidori,
 JHEP 09 (1998) 009 [hep-ph/9808487].}%
\nref\NiSiZ{Y. Nir and D. Silverman,
 Phys. Rev. D42 (1990) 1477.}%
\nref\BMPR{G.C. Branco, T. Morozumi, P.A. Parada
 and M.N. Rebelo, Phys. Rev. D48 (1993) 1167.}%
\nref\Poma{A. Pomarol,
 Phys. Rev. D47 (1993) 273 [hep-ph/9208205].}%
\nref\BaBa{K.S. Babu and S.M. Barr,
 Phys. Rev. Lett. 72 (1994) 2831 [hep-ph/9309249].}%
\nref\AbFr{S.A. Abel and J.M. Frere,
 Phys. Rev. D55 (1997) 1623 [hep-ph/9608251].}%
\nref\EyNiacp{G. Eyal and Y. Nir,
 Nucl. Phys. B528 (1998) 21 [hep-ph/9801411].}%
\newsec{Conclusions}
A low value of $\apks$, say $\apks\lsim0.5$, is well within
the average of present measurements, $\apks=0.42\pm0.24$.
It is outside however of the indirect constraint on $\sin2\beta$ 
that holds within the Standard Model. We discussed two general 
scenarios in which the intriguing possibility of a low $\apks$
can be accommodated:

1. The Standard Model is valid. The constraints on $\sin2\beta$
are wrong because one or more of the relevant hadronic parameters
are outside of the range that is considered in the literature.
The possible explanations are as follows:
\item{(i)} $|V_{ub}|$ is smaller than its theoretically favored range.
\item{(ii)} $\xi$ is larger than its theoretically favored range.
\item{(iii)} $\hat B_K$ is larger than its theoretically favored range.
\item{(iv)} At least two of the theoretical parameters are
outside of the preferred ranges. 

2. Present estimates of all hadronic parameters are correct.
There is new flavor violating and/or CP violating physics.
The possible scenarios of new physics are as follows:
\item{(i)} There are significant new contributions to $\BBbar$ mixing
and to $\BBsbar$ mixing. They can carry the SM phase but 
in such a case their flavor violation is different from the CKM one.
\item{(ii)} There is a significant new CP violating contribution to
$\BBbar$ mixing. 
\item{(iii)} There is a new CP violating contribution to $\KKbar$
mixing and very likely also to $K\ra\pi\pi$ decays. 
\item{(iv)} All of the above effects can simultaneously play a role
in modifying both the relation between $\apks$ and $\sin2\beta$
and the SM predictions for $\sin2\beta$.
\item{(v)} In addition, if there are extra quarks beyond the three
generations of the SM, there are more ways in which the CKM constraints
can be modified \EyNi. However, in such models, the dominant effect
is always a new contribution to the mixing \NiSi.

It is possibe now to consider specific models of new physics
and examine whether they can lead to small values of $\apks$.
For example, within the supersymmetric framework, many models
that do not have exact universality can significantly modify
both $\BBbar$ mixing and $\KKbar$ mixing (for a review, see \gnr),
thus allowing for a small $\apks$. On the other hand,
in models of exact universality such effects are generically 
small \dgh. Another interesting example is that of models with extra, 
SU(2)-singlet down quarks. Here, the new contribution to
$\epsK$ is small \refs{\BuSi,\CoIs}\ but the relation
between $\apks$ and $\sin2\beta$ can be significantly modified
\refs{\NiSiZ,\BMPR}, allowing for low $\apks$ values.

Finally, we would like to note that the allowed range for
$\apks$, eq. \aveapks, is consistent with zero asymmetry
at the 1.75$\sigma$ level and certainly does not exclude
the possibility that the asymmetry is small. This leaves 
viable a framework that is drastically different from the
Standard Model where CP is an approximate symmetry of the full 
theory, that is, CP violating phases are all small 
\refs{\Poma-\EyNiacp}. 

\vskip 0.5cm

\centerline{\bf Acknowledgments}
We thank Joao Silva for useful discussions.
Y.N. is supported by the Israel Science Foundation founded by the 
Israel Academy of Sciences and Humanities, by the United States $-$ 
Israel Binational Science Foundation (BSF) and by the Minerva 
Foundation (Munich).  

\listrefs
\end